\documentclass[prb,preprint]{revtex4}
\pdfoutput=1
\usepackage{graphicx}
\begin{document}


\title{Is stripe incommensurability in $\mathbf{La_{2-x}Sr_xCoO_4}$ proportional to doping?
\medskip }

\date{October 15, 2017} \bigskip

\author{Manfred Bucher \\}
\affiliation{\text{\textnormal{Physics Department, California State University,}} \textnormal{Fresno,}
\textnormal{Fresno, California 93740-8031} \\}

\begin{abstract}
The longstanding notion of stripe incommensurability being proportional to doping, $\delta(x) \propto x$, in lanthanum transition-metal oxides, $La_{2-x}Sr_{x}TmO_4$ ($Tm = Cu, Ni, Co$), is partly borne out by experiment but also plagued with exceptions. Future neutron-scattering experiments on cobaltates could provide a clear distinction whether a linear or square-root dependence, $\delta(x) \propto \sqrt{x - x_0}$,\, is valid.

\end{abstract}

\maketitle

Pristine lanthanum cuprate, nickelate and cobaltate, $La_2CuO_4$, $La_2NiO_4$ and $La_2CoO_4$, are antiferromagnetic Mott insulators of the same `214' crystal type. When doped with divalent $Sr$---or likewise with $Ba$ in $La_2CuO_4$---ionized lanthanum atoms, $La \rightarrow La^{3+} + 3e^-$, are substituted by ionized $Sr \rightarrow Sr^{2+} + 2e^-$. This causes electron deficiency (hole doping) of concentration $n_h = x$. The doped holes destroy magnetic dipole moments in the host crystal (by a still unknown mechanism), thereby weakening the magnetic interaction. Three-dimensional antiferromagnetism (3D-AFM) breaks down with hole doping at the N\'eel concentration, $x^N(T)$, with values of $x_0^N = 0.02, \, \simeq 0.11$ and $\approx 0.25$ at temperature $T=0$ in the respective compounds. The increasing stability of the corresponding 3D-AFM is a consequence of larger spin magnetism with $S=\frac{1}{2}$, $S=1$ and $S=\frac{3}{2}$ of the $Cu^{2+}$ $3d^9$, \, $Ni^{2+}$ $3d^8$ and $Co^{2+}$ $3d^7$ ions, respectively. With more $Sr$ doping 2D-AFM persists and magnetic density waves (MDWs) appear---in case of the cuprates together with charge-density waves (CDWs). The density waves are incommensurable with the crystal lattice and therefore often called ``stripes.'' Interest in stripes arises from their possible connection with superconductivity in the cuprates and with the enigmatic pseudogap phase.\cite{1} Qualitatively the incommensurability $\delta$ of the stripes increases with doping $x$. Early experiments with neutron scattering\cite{2} indicated the incommensurability of MDWs in $La_{2-x}Sr_xCuO_4$ to be proportional to the level of hole doping
(see Fig.1 at $0.03 \leq x \leq 0.05$),
\begin{equation}
\delta(x) = c \, x \,, 
\end{equation}
where $c$ is a coefficient of proportionality.
Later experiments with neutron or X-ray scattering in extended doping ranges of $La_{2-x}Sr_xCuO_4$ showed that the incommensurability of MDWs and CDWs deviates from proportionality, Eq. (1), leveling off for $x > 1/8$.

A derivation of the doping dependence of stripe incommensurability,\cite{3} based on a partition of the $CuO_2$ plane by the doped holes gave a square-root dependence,
\begin{equation}
\delta(x) = C \sqrt{x - x_0} \, , 
\end{equation}
with an off-set by the N\'eel concentration, $x_0 = x_0^N$, in the radicand. The coefficients $c$ and $C$ in Eq. (1) and Eq. (2) depend on the orientation of the stripes (parallel or diagonal to the transition-metal oxygen bond, $Tm$-$O$ ($Tm = Cu, Ni, Co$), and on the kind of stripes (MDWs or CDWs). Measured incommensurabilitys of stripes in
cuprates 
$La_{2-x}Ae_xCuO_4$ \, ($Ae = Sr, Ba)$,
nickelates 
$La_{2-x}Sr_xNiO_4$ and
cobaltates 
$La_{2-x}Sr_xCoO_4$, together with graphs of Eq. (1) (hatched diagonal) and Eq. (2) (solid curve) are shown in Figs. 1 - 3.

\includegraphics[width=5.98in]{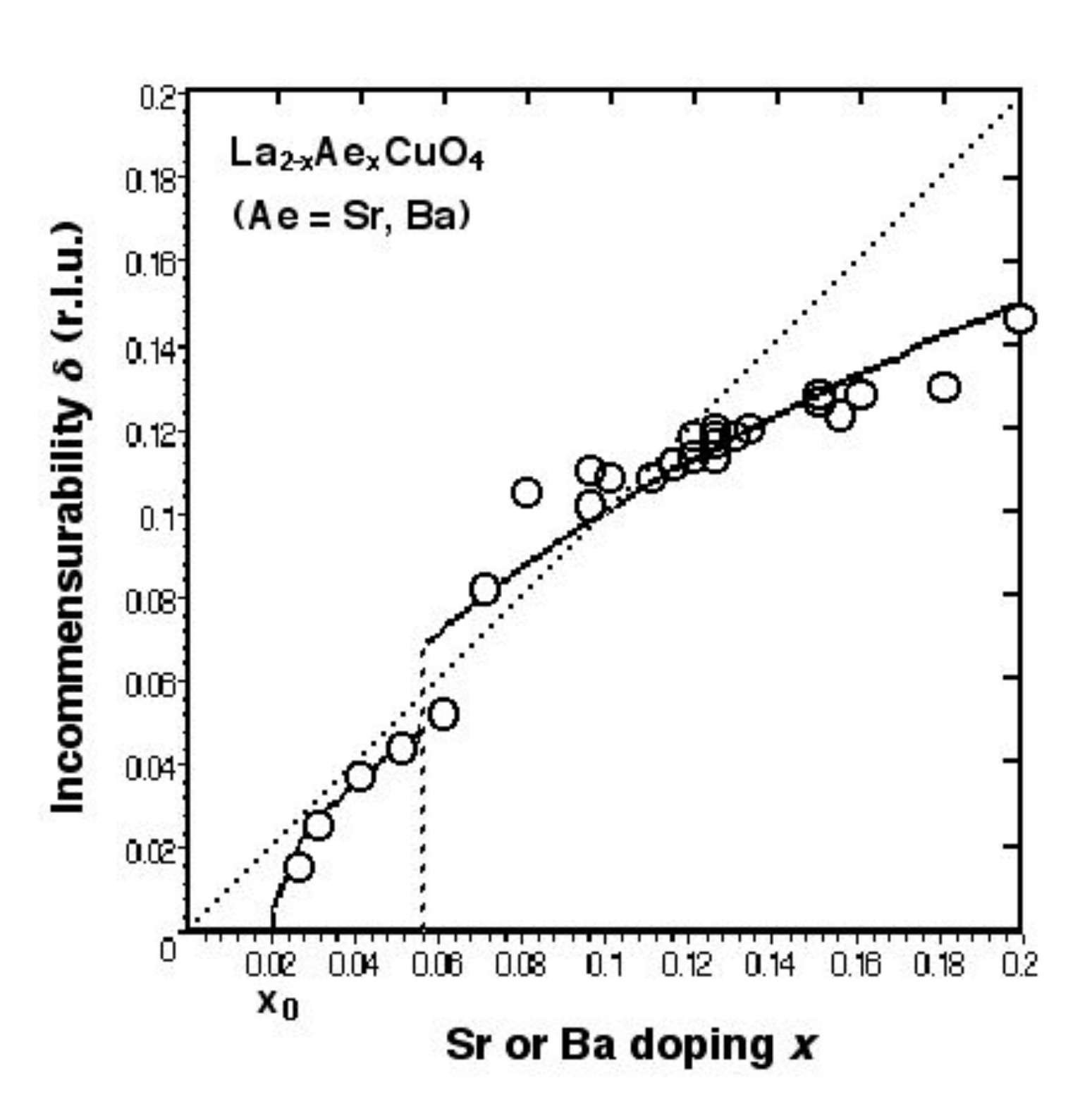}

\footnotesize
\noindent FIG. 1. Incommensurability of stripes in $La_{2-x}Ae_{x}CuO_{4}$ due to doping with $Ae = Sr$ or $Ba$. Stripe orientation is diagonal for $x < 0.056$ and parallel to the $Cu$-$O$ bonds for $x > 0.056$, giving rise to a jump of $\delta(x)$ at $x = 2/6^2 \simeq 0.056$. Circles show data from neutron scattering or X-ray diffraction cited in Ref. 4. The broken solid curve is a graph of Eq. (2), taking into account the change of stripe orientation at $x \simeq 0.056$ and the relation $\delta(x) = \delta_m(x) = \frac{1}{2}\delta_c(x)$ between magnetic ($m$) stripes and charge-density ($c$) stripes. The dotted diagonal line shows the proportionality $\delta(x) = x$, Eq. (1).
\normalsize

The data in Fig. 1 fall near the diagonal for low doping, $0.03 \leq x \leq 0.05$, as mentioned, and again near the intersection of the curve and diagonal at $\dot{x} = 0.10$. But they clearly deviate from the diagonal otherwise, particularly for large doping levels, $0.12 < x \leq 0.20$. 
Concerning the latter deviation it is not clear whether or not the data ``level off'' (ramp-like), as frequently noted in the literature, implying a \emph{constant level} of $\delta(x)$ at large $x$.

\includegraphics[width=6in]{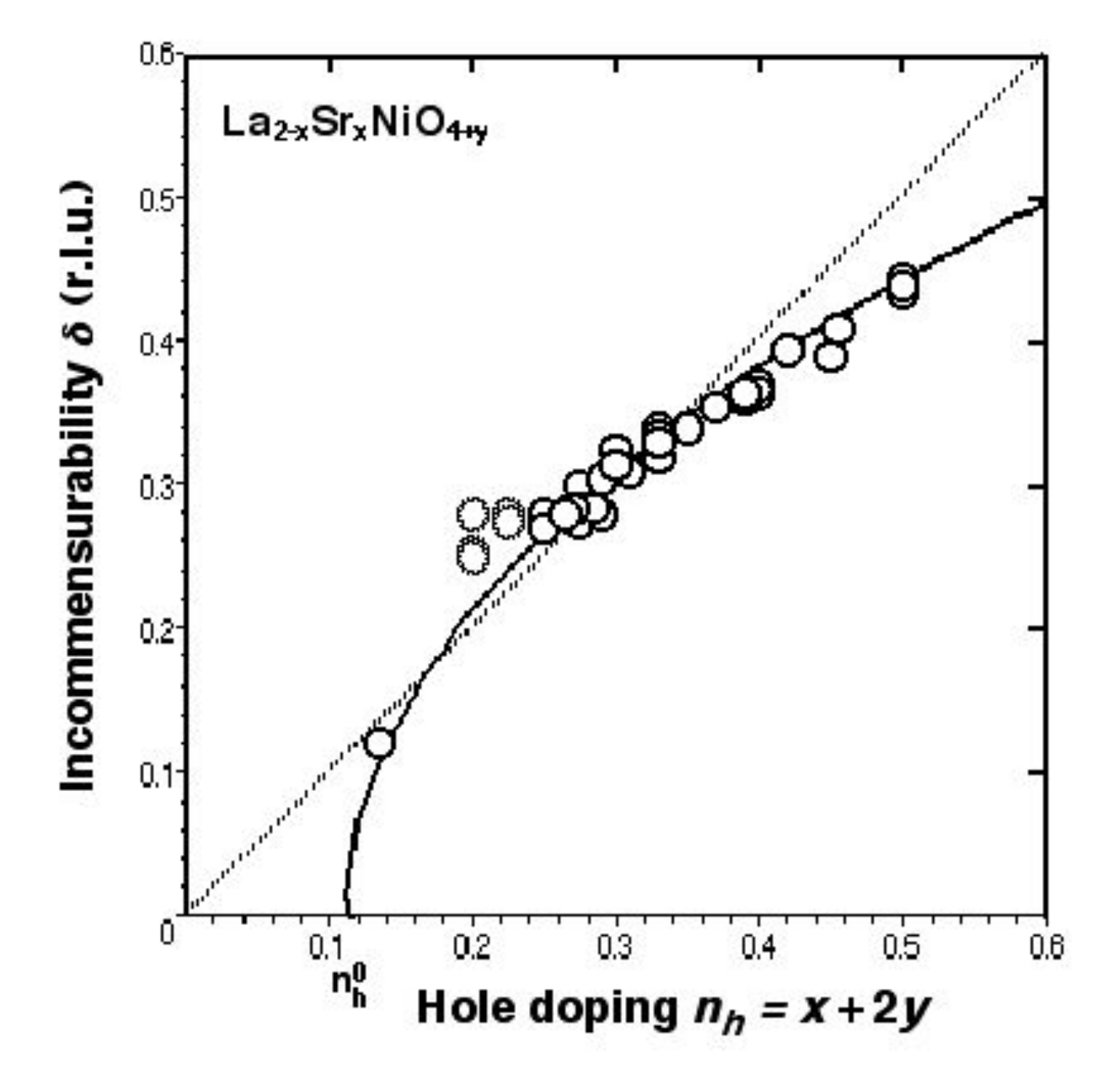}
\footnotesize
\noindent FIG. 2. Incommensurability $\delta(n_h)$ of stripes in $La_{2-x}Sr_{x}NiO_{4+y}$ due to hole doping by $n_h = x +2y$. Circles show data from neutron scattering or X-ray diffraction cited in Ref 5. The solid curve is a graph of Eq. (2). The hatched diagonal line depicts the proportionality $\delta(n_h) = n_h$, Eq. (1). Mottled circles in the doping range $0.20 \leq n_h < 0.25$ indicate very low stability of stripes.
\normalsize
\bigskip \bigskip 

The data in Fig. 2 fall near the diagonal line around the high intersection of the curve and diagonal at $\dot{x} = 1/3 \simeq 0.333$ where stripes have the highest stability. 
However, the data clearly deviate from the diagonal for high doping, $0.4 \leq n_h \leq 0.5$.

\includegraphics[width=6in]{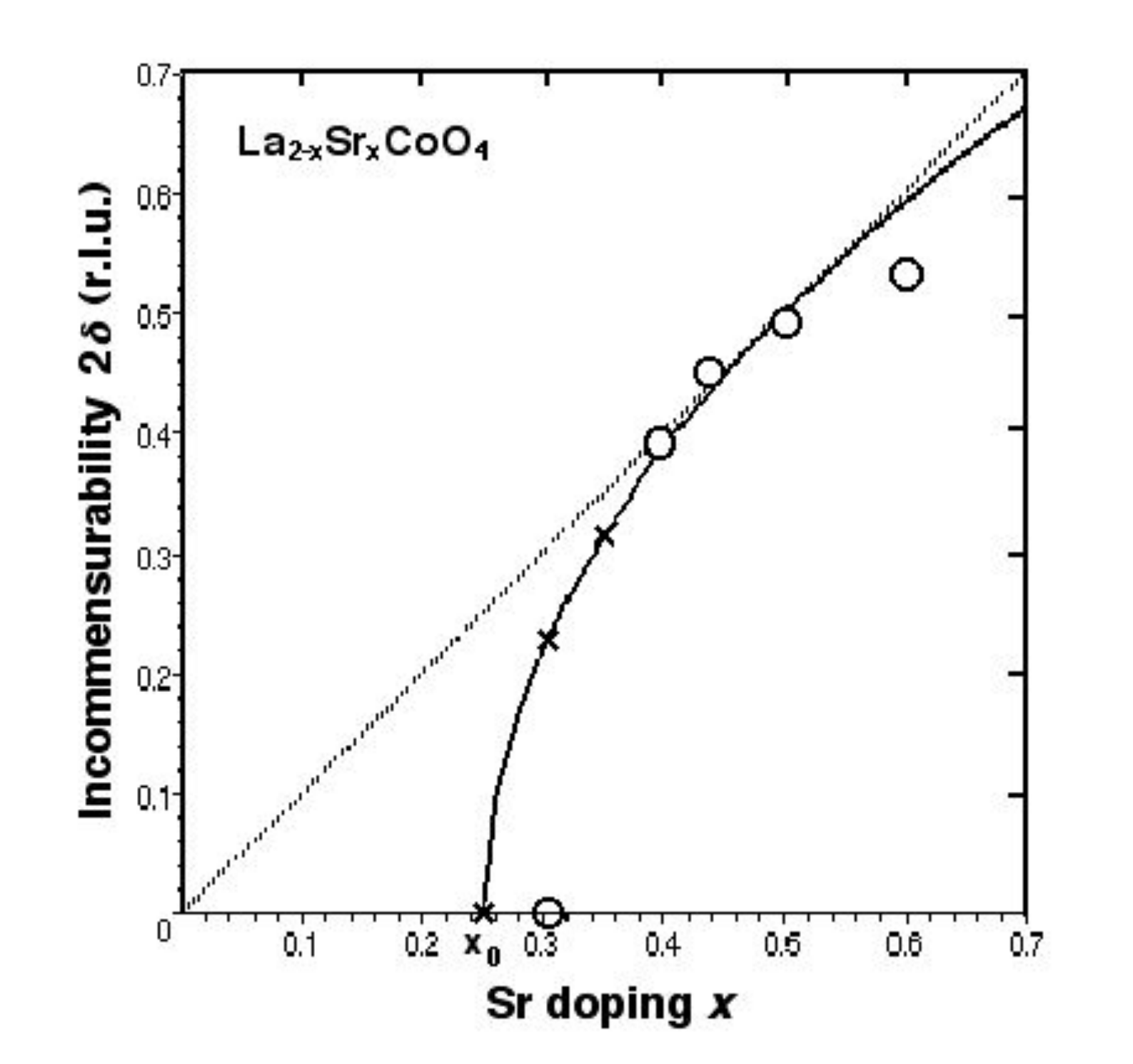}

\footnotesize 
\noindent FIG. 3. Doubled incommensurability, $2 \delta$, of magnetic stripes in $La_{2-x}Sr_{x}CoO_4$ in dependence on $Sr$ doping $x$. Circles show experimental data from neutron scattering [Ref. 6]. The solid curve is a graph of Eq. (2) with coefficient $C = 1/2$. The hatched diagonal displays proportionality with doping, $2 \delta(x) = x$, Eq. (1). Cross marks ($\times$) show theoretical values used for extrapolated spectra in Fig. 4.
\normalsize
\bigskip

Three of the data in Fig. 3---for doping $0.4 \leq x  \leq 0.5$ and near the tangent point at $\dot{x} = 0.5$---fall equally close to the curve and diagonal. The deviating case at $x=0.6$ may need special consideration as it falls into the checkerboard pattern starting at half-doping, $x=0.5$.\cite{7} Absence of stripe incommensuration, $\delta = 0$,  was observed for $x = 0.3$ by commensurate magnetic scattering and confirmed by commensurate AFM character of spin-wave modes.\cite{6} 

\pagebreak

Comparing Figs. 1 -  3, the incommensurability $\delta(x)$ appears linear in the doping ranges where the diagonal line and square-root curve of Eqs. (1) and (2) are close. The situation is somewhat peculiar in the case of the cuprates, Fig. (1), where---because of the jump at $x \simeq 0.056$---the broken square-root curve stradddles the diagonal for a long stretch. Nevertheless, a square-root dependence of $\delta(x)$ can be seen both in the low-doping range of diagonal stripes and in the high-doping range of axially parallel stripes. 
In the nickelates no data are available in the low-doping range (except for $x = 0.135$), in part due to stripe instability at the lower diagonal-curve intersection at $x = 1/6 \simeq 0.167$.\cite{5} Otherwise such data could strengthen (or weaken) the claim that a square-root dependence, Eq. (2), represents stripe incommensurability $\delta(x)$ better than proportionality, Eq. (1). However, for the cobaltates a discrimination may be achieved if signals could be measured in the unexplored doping range, $0.3 < x < 0.4$, when neutron-scattering experiments are revisited. From the values of $\delta(0.3)$ and $\delta(0.4)$ the doubled incommensurability of the corresponding stripes can be expected to fall below or at the diagonal, $2\delta(x) \leq x$. 
 
 Experimentally, the doubled incommensurabily of magnetic stripes in the cobaltate $La_{2-x}Sr_{x}CoO_4$ was determined (for $x > 0.3)$ by the separation of double peaks in the neutron scattering spectrum.\cite{6} Simplifying the situation with Gaussian peaks and ignoring background, the measured scattering spectra are reproduced from Ref. 6 in Fig. 4 for $x = 0.50,\, 0.45,\, 0.40$. As can be seen, with lesser doping the separation of the double peaks decreases while the width of each peak increases dramatically. We now want to extrapolate the measured spectra to lower doping levels, $x = 0.35,\, 0.30$, $0.25$ (indicated by cross marks in Fig. 3), based on the assumptions (i) that the square-root formula for the incommensurability is valid---Eq. (2) with a coefficient $C = 1/2$---and (ii) that the trend of widening peak-width continues progressively. The goal is not high accuracy (impeded by inaccuracy of peak widening) but qualitative insight.
 
 As a consequence of double-peak approach and the dominating effect of peak widening, the double peaks will eventually start to overlap, causing a rise of their inner flanks. This can be seen qualitatively in Fig. 4 for $x = 0.35$. The incommensurability would still be measurable by the double-peak separation. The situation changes, however, when the sum of the double peaks forms a \emph{single} peak as qualitatively shown in Fig. 4 for $x = 0.30$. Now the separation of the constituting double peaks, $2\delta$, can no longer be determined from the measurable single-peak signal. Instead, $2\delta$ must be determined by line-shape analysis. As 
 
 \includegraphics[width=6in]{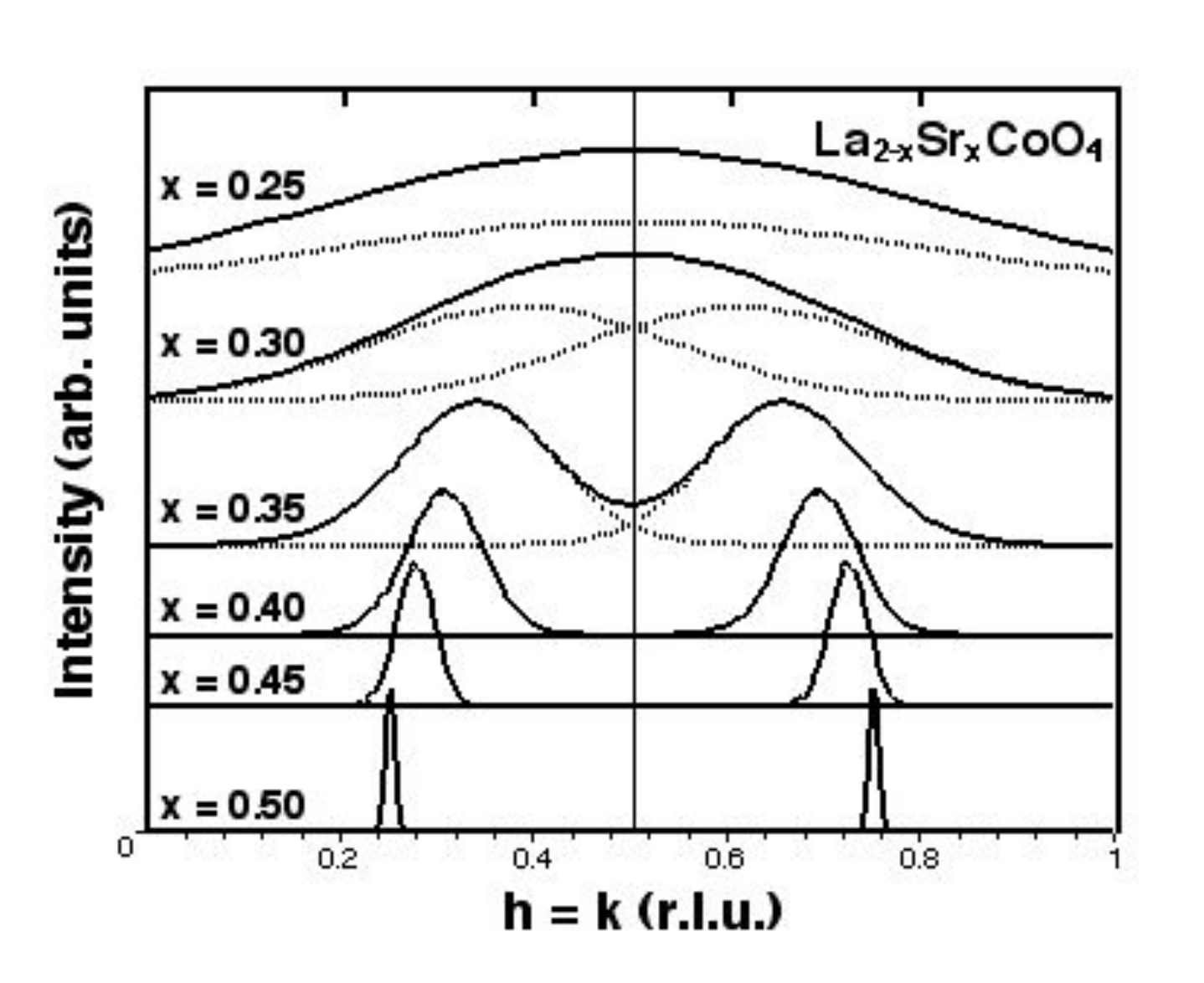}

\footnotesize 
\noindent FIG. 4. Elastic magnetic scattering in $La_{2-x}Sr_{x}CoO_4$ (simplified by Gaussian profiles and without background signal, vertically displaced) from measurements ($x = 0.50, 0.45, 0.40$ [Ref. 6]), scanned in the $(h,k,0)$ plane, and extrapolated to $x = 0.35, 0.30, 0.25$ with peak positions from Eq. (2) but extended trend of peak width. Solid lines represent the measurable signal, hatched lines show constituting double peaks (coinciding for $x = 0.25$).
\normalsize
\bigskip

\noindent the case illustrates, the appearance of a single peak in the measured signal is a necessary, but \emph{not} a \emph{sufficient} condition for the disappearance of stripe incommensurability, $\delta = 0$.

In the final scenario, $x = 0.25$, the constituting double peaks coincide. Only in this case can $\delta = 0$ be concluded from the single-peak shape of the measured signal. These considerations may be useful in future neutron-scattering experiments on $La_{2-x}Sr_xCoO_4$ with the goal of discriminating between linear or square-root dependence of stripe incommensurability $\delta(x)$, Eq. (1) or (2).
\pagebreak

\centerline{ \textbf{ACKNOWLEDGMENTS}}
\noindent I thank Duane Siemens for valuable discussions and Preston Jones for help with graphics.

\end{document}